\documentclass[12pt]{article}

\usepackage{amssymb,esvect,amsmath,graphicx,latexsym,amsthm,slashed,eso-pic,hyperref}
\usepackage[final]{pdfpages}

\usepackage{epsfig,amsmath,amssymb,verbatim,mathrsfs,hyperref}
\usepackage{xspace}
\usepackage{xcolor}

\usepackage{slashed}
\usepackage{dcolumn}
\usepackage{multirow}
\graphicspath{{./}{Figs/}}
\usepackage{feynmp-auto}

\def\beq{\begin{eqnarray}}
\def\eeq{\end{eqnarray}}
\def\bea{\begin{eqnarray}}
\def\eea{\end{eqnarray}}

\newcommand{\gsim}{\lower.7ex\hbox{$\;\stackrel{\textstyle>}{\sim}\;$}}
\newcommand{\lsim}{\lower.7ex\hbox{$\;\stackrel{\textstyle<}{\sim}\;$}}

\begin{document}
\begin{titlepage}
\begin{flushright}

\end{flushright}

\vskip 2.2cm

\begin{center}

{\Large \bf Precision Inclusive Higgs Physics at $e^+e^-$ Colliders \\
\vspace{.3cm} With Tracking Detectors and Without Calorimetry }

\vskip 1.4cm

{\large Patrick Draper$^{(a)}$, Jonathan Kozaczuk$^{(a,b)}$, and Scott Thomas$^{(c)}$ }
\\
\vskip 1cm
{\small 
$^{(a)}$ Department of Physics, University of Illinois, Urbana, IL 61801\\
$^{(b)}$ Amherst Center for Fundamental Interactions, Department of Physics,\\ University of Massachusetts, Amherst, MA 01003\\
$^{(c)}$ New High Energy Theory Center, Rutgers University, Piscataway, NJ 08854, USA
}\\
\vspace{0.3cm}
\vskip 4pt

\vskip 1.5cm

\begin{abstract}

\noindent A primary goal of a future $e^+e^-$ collider program will be 
the precision measurement of Higgs boson properties.   
For practical reasons it is of interest to determine the minimal set of detector specifications
required to reach this and other scientific goals.   
Such information
could be useful in developing a staged approach to the full collider project with 
an initial lower-cost version focused on achieving some of the primary scientific objectives. 
Here we investigate the precision obtainable for the   
$e^+ e^- \rightarrow Z h \rightarrow \mu^+ \mu^- X$ inclusive cross section and the Higgs boson mass 
using the di-muon recoil method,
considering a detector that has only an inner tracking system within a solenoidal magnetic field,
surrounded by many nuclear interaction lengths of absorbing material, and an outer muon identification system.
 We find that 
the sensitivity achievable in these measurements with such a tracking detector is only marginally 
reduced compared to that expected for a general purpose detector with additional electromagnetic and hadronic calorimeter systems.
The difference results mainly from multi-photon backgrounds that are not as easily rejected with tracking detectors.
We also comment on the prospects for an analogous measurement of the inclusive cross section $\sigma( e^+ e^- \to Z h \to e^+ e^- X)$. 
Finally, we study searches for light scalars utilizing the di-muon recoil method, estimating the projected reach with 
a tracking and general purpose detector.

\end{abstract}

\end{center}

\vskip 1.0 cm

\end{titlepage}

\setcounter{footnote}{0} 
\setcounter{page}{1}
\setcounter{section}{0} \setcounter{subsection}{0}
\setcounter{subsubsection}{0}
\setcounter{figure}{0}

\section{Introduction}


The discovery of the 125 GeV Higgs boson at the LHC was a milestone achievement in high energy physics, completing the Standard Model (SM) and validating the weakly-coupled Higgs mechanism as the source of electroweak symmetry breaking. Within current measurement precision, all observed Higgs production and decay rates have been in agreement with SM predictions. The long-term LHC program is expected to measure many of the Higgs couplings at the 5-10\% level (exclusive Higgs rates at the 10-20\% level).

Looking further into the future, it is highly desirable to put the Higgs boson under a microscope, obtaining qualitatively new measurements of its properties and improving the experimental precision in currently measured rates to the next order of magnitude. Proposed $e^+ e^-$ colliders including the ILC, CEPC, FCC-ee, and CLIC, would provide extremely clean, high-statistics samples of Higgs bosons, offering extensive discovery potential for new physics beyond the SM connected with the Higgs. The physics case for these colliders is now broadly established~\cite{ilctdr, ilc250, cepcprecdr,cepccdr, tlepwg,clic} and they are at various stages of design. It remains to be seen which, if any, will be constructed. 

\begin{figure}[t!]
\begin{center}
\includegraphics[width=.8\linewidth]{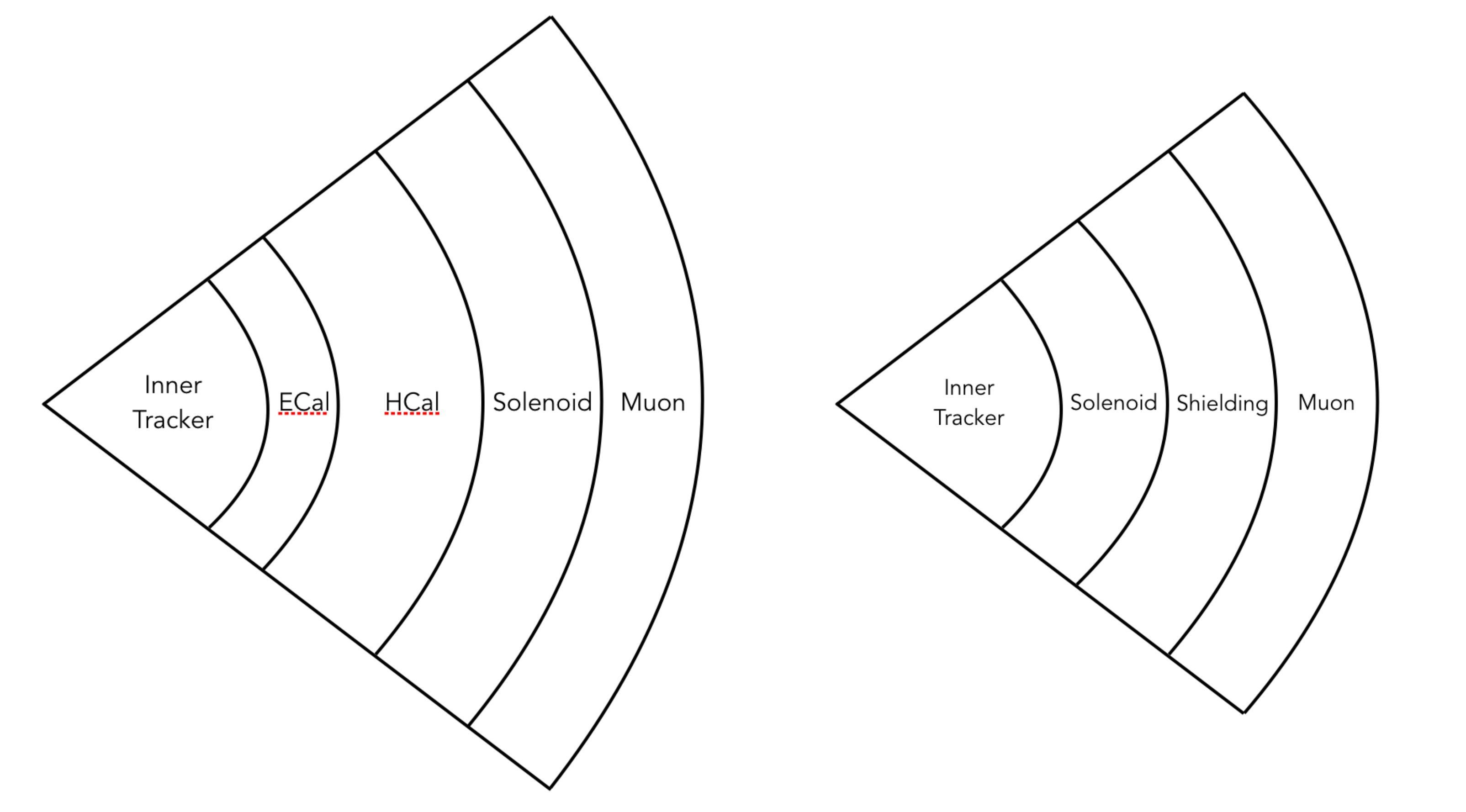} 
\caption{Cartoon comparing a transverse wedge slice of a general purpose collider detector (left) 
with a similar slice of a magnetic solenoid tracking detector (right). 
The inner tracker systems are similar in both detectors. 
The general purpose detector has electromagnetic and hadronic calorimeter systems 
outside the inner tracker. 
The magnitude of the solenoidal magnetic field in the inner tracker region  
is similar in both detectors. 
However, the solenoid can be physically smaller in the tracking detector.  
In the general purpose detector the return yoke for the solenoid is embedded within the 
muon system. 
In the tracker detector the return yoke can be part or all of the shielding material outside 
the solenoid.} 
\label{fig:detector}
\end{center}
\end{figure}

It is interesting to ask what is the minimal Higgs factory capable of achieving key science goals. This question may  be of practical use, particularly if it would be logistically or financially advantageous to consider a staged approach to building the detectors. Staging, for example, could smooth out up-front costs over a period of several years. In this work we study one of the most important Higgs measurements at $e^+ e^-$ colliders, the inclusive $\sigma_{Zh}$ cross section measurement based on the $Z$ recoil method, using an analysis relying only on tracking and muon system data, without calorimetry.  Comparing to a traditional analysis with full calorimetry, we find that the track-based analysis is nearly as sensitive in the muon channel, respectively reaching 6\% and 5\% expected precision with about one or two years of data. The Higgs boson mass can also be measured in this channel, for which again we find only a marginal reduction in sensitivity for tracking detectors without calorimeters. A cartoon of the two detectors we consider in the analysis is shown in Fig.~\ref{fig:detector}. The small difference in reach for these measurements arises primarily from multiphoton backgrounds, which are more easily vetoed with electromagnetic calorimetry. 

Before continuing on, perhaps it is worth elaborating further on our motivation for this study. We are not necessarily advocating for building a stripped-down version of a detector in place of the established concepts already under consideration. Instead, allowing for certain costly subsystems to be added at different points in the operational timeline can provide the flexibility required to satisfy practical constraints (e.g.~annual budgetary limitations) that are likely to arise during the course of a realistic experiment. While this may seem like a secondary issue, it is important to consider the possibility of a staged approach now, as it could impact the fundamental design of the individual components which are currently being discussed for the various proposed experiments. Our line of reasoning shares some similarities with the motivation behind the Fourth Concept Detector~\cite{fourthconcept} proposed for the ILC, where simplicity and a reduction in the number of detector components was a guiding design principle, and the IDEA detector~\cite{cepccdr} for circular colliders, which aims to reduce the overall cost of the detector. Besides its pragmatic advantages, a staged approach to a Higgs factory would provide opportunities for creativity and innovation in detector design that we hope might be of interest to members of the experimental community.

This work is organized as follows. In Sec.~\ref{sec:recoil} we review the $\sigma_{Zh}$ measurement and describe our signal and background generation, commenting on several related technical issues. In Sec.~\ref{sec:analysis} we perform the track-based analysis and compare it to a full calorimetric analysis, also commenting on sensitivity to the Higgs mass and simple alternative methods for photon rejection. In Sec.~\ref{sec:electrons} we discuss the $\sigma_{Zh}$ measurement in the electron channel, where backgrounds are larger, final state radiation (FSR) and bremsstrahlung are more important, and discriminating pions from electrons is essential. We do not perform a complete analysis here, but we show that hadronic backgrounds can be efficiently rejected with tracking information alone. In Sec.~\ref{sec:light_scalar} we perform a closely-related analysis of searches for new light bosons mixing with the Higgs using the $Z\to\mu^+\mu^-$ recoil method with and without calorimetry. In Sec.~\ref{sec:discussion} we summarize and conclude, commenting on directions for future study.

\section{$e^+e^-\to Zh\to\mu^+\mu^-+X$} 
\label{sec:recoil}
One of the most important measurements expected to be performed at future $e^+ e^-$ colliders is that of the $Zh$ production cross-section, $\sigma_{Zh}$. The coupling of the Higgs to $Z$ bosons is very sensitive to new physics beyond the Standard Model. For example, $\sigma_{Zh}$ can deviate significantly from its predicted SM value if the Higgs mixes with another scalar, and is generally affected if the Higgs couples to new states through the ``Higgs portal'' mechanism.  Deviations in $\sigma_{Zh}$ can be tied to solutions to the hierarchy problem, dark matter, or new physics predicting a strong first-order electroweak phase transition~\cite{Craig:2013xia, Craig:2014una, AboveThreshold, Curtin:2014jma, Assamagan:2016azc, Huang:2016cjm, Chen:2017qcz}. Most proposals for future $e^+ e^-$ colliders ultimately aim to achieve percent-level precision in $\sigma_{Zh}$~\cite{ilctdr,cepcprecdr, tlepwg, LEP3, CEPC_Higgs,  ILC_Higgs, CLIC_Higgs}, which would provide unparalleled sensitivity to new physics involving the Higgs. It is important that any detector design and experimental run plan allow for this level of sensitivity in order to maintain the physics case for these colliders.

The excellent sensitivity to $\sigma_{Zh}$ achievable at $e^+ e^-$ experiments is due to the relatively clean environment and the ability to perform a recoil measurement, independent of the Higgs decay mode, to isolate genuine $Zh$ events. Typically, the cleanest channel is $Z(\to \mu^+ \mu^-)h$, as muon identification is expected to be very efficient and the corresponding momenta can be measured with high accuracy. The electron channel $Z(\to e^+ e^-)h$ is typically expected to offer slightly less sensitivity, due to additional backgrounds and the effects of final state radiation and bremsstrahlung. The hadronic channel $Z(\to \overline{q} q)h$ provides larger statistics, but suffers larger backgrounds and less resolution. Here, we study the prospects for measuring $\sigma_{Zh}$ in the muon channel with and without detector calorimetry. As we review below, such a measurement typically relies primarily on the kinematics of the muons, which can be accurately determined with tracking alone. 

\subsection{Signal}

For the signal, we consider the process
\beq
e^+ e^- \to Z h, \qquad Z \to \mu^+ \mu^-, \, h\to {\rm anything}.
\eeq
 Measuring the rate for this process and dividing by the well-measured $Z\to \mu^+\mu^-$ branching ratio yields a direct measurement of $\sigma_{Zh}$. To determine the ultimate sensitivity to $\sigma_{Zh}$ in this channel, we will compare the predicted signal and corresponding backgrounds. We generate $e^+ e^- \to \mu^+ \mu^- h$ events using the Monte Carlo event generator \texttt{Whizard 2}~\cite{whizard_1, whizard_2}, including the effects of initial state radiation (ISR). The effects of beamstrahlung are not included, as they are expected to be small at circular $e^+ e^-$ colliders. For the Monte Carlo samples we analyze below, only signal events with prompt muons are included, as the contribution from events with muons arising from heavy flavor decays to the signal region is found to be negligibly small. Throughout our study the center-of-mass energy $\sqrt{s}$ is taken to be $\sqrt{s}=250$ GeV. Events are passed to \texttt{Pythia 6}~\cite{pythia} for final state showering and hadronization. At the generator level, $\sigma_{Zh} \times BR(Z \to \mu^+ \mu^-) \simeq 7$ fb in the Standard Model.  Details concerning the reconstruction and selection criteria are described below in Sec.~\ref{sec:recoil_details}. All signal and background samples are generated assuming 200 fb$^{-1}$ of integrated luminosity, which could conservatively be achieved within one or two years of running~\cite{ilc250,cepccdr, tlepwg}.

\subsection{Backgrounds}
In order to determine the precision with which one can measure $\sigma_{Zh}$, several background processes mimicking the signal must also be accounted for.  The dominant contributions can be grouped into two fermion ($2f$)  and four fermion ($4f$) processes, described below. In our analysis we again utilize \texttt{Whizard 2} and \texttt{Pythia 6} for background event generation and showering/hadronization. 

\subsubsection{$2f$+ISR}
An important background for the $Zh$ recoil analysis consists of two-fermion events, primarily produced either through a $Z$ or off-shell photon, accompanied by a significant amount of initial state radiation. The ISR photons can provide the necessary ``kick'' for the recoil mass, $m_{\rm reco}$ (defined below), to fall in the Higgs mass window. While our analysis will include cuts on the lepton $p_T$, the reconstructed $Z$ mass, and $m_{\rm reco}$, it is straightforward to see that $2f$ events with at least one sufficiently transverse photon and one hard collinear photon can contribute to the signal region. Thus, low-$p_T$ hard ISR photons should be accounted for in our analysis. 

Emission of $n$ hard, collinear ISR photons results in logarithmically-enhanced contributions to the inclusive $2f$ cross-section proportional to
\beq \label{eq:log}
\frac{1}{n!} \alpha_{\rm EM}^n\log^n \frac{s}{m_e^2}.
\eeq
These contributions can be accounted for, up to a given $n$, by electron structure functions in \texttt{Whizard}. Although these structure functions marginalize over the ISR $p_T$,  \texttt{Whizard} can also generate events with finite transverse momentum for the recoiling system so that the distributions reproduce the expected behavior in $p_T$. We utilize this feature when generating events. Since the two fermions produced by \texttt{Whizard} will be required to yield muons reconstructing to near $m_Z$, we incorporate a generator level cut on the fermion-antifermion invariant mass, $m(f\overline f) > 10$ GeV. After imposing our selection criteria, we find that the dominant contribution is from events with two prompt muons\footnote{In what follows, we use ``muons'' to refer to both $\mu^+$ and $\mu^-$, except where noted.}, $e^+ e^- \to \mu^+ \mu^- + {\rm ISR}$, and so we neglect the non-prompt $2\mu$ background in what follows.
 
In comparing detectors with and without an electromagnetic calorimeter (ECAL), we will be interested in the effect of a photon veto on this background, since some sufficiently hard transverse ISR is required for $2\mu$ processes to contribute to the signal region. Events with a single hard photon will have $m_{\rm reco}=0$ at parton-level, and thus not contribute to the signal region. Meanwhile, for $\sqrt{s}=250$ GeV, the effective expansion parameter in Eq.~(\ref{eq:log}) is not very large, $\alpha_{\rm EM}\log s/m_e^2 \sim 0.2$, and so the contribution of hard $3 \gamma$ events is suppressed by $\sim \mathcal{O}(0.1)$ relative to events with two hard photons. We therefore expect the largest background contribution from  $e^+ e^- \to \mu^+ \mu^-+ 2 \gamma$. For each event, the \texttt{Whizard} approach effectively generates one ISR photon per beam with non-zero $p_T$, and so should reasonably capture the transverse photon kinematics in the signal region. Note that \texttt{Madgraph}~\cite{madgraph} can also be used to analyze this background, provided that one includes two photons in the generated process, incorporates non-zero lepton masses (to cut off collinear divergences), and sets a low generator-level photon energy cut.

\subsubsection{$4f$}

Four-fermion processes ($e^+ e^- \to 4 f$) make up another important background for the recoil measurement when the final state involves a $\mu^+ \mu^-$ pair, produced directly in the hard process and/or in heavy flavor decays. In the latter case, the muons tend to be produced in close proximity to additional charged tracks with significant $p_T$. After imposing muon isolation criteria, we find that events with two non-prompt muons contribute negligibly to this background, and so we consider only $e ^+ e^- \to  \mu^+ \mu^- + 2 f$ and $e ^+ e^- \to \mu^\pm + 3 f ( \neq \mu$) in what follows. 

\begin{figure}[!t]
\centering
\includegraphics[trim=10 0 0 0,clip,width=.3\textwidth]{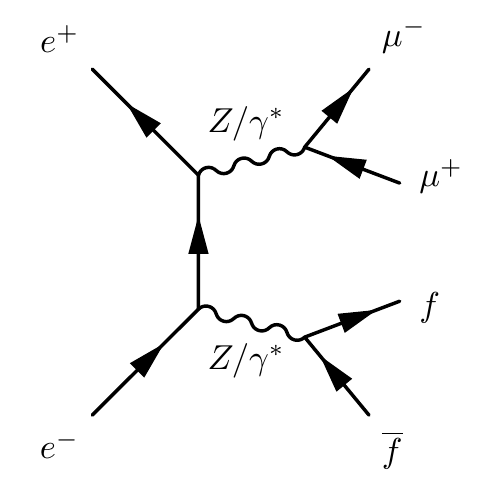} \quad  \includegraphics[trim=10 0 0 0,clip, width=.3\textwidth]{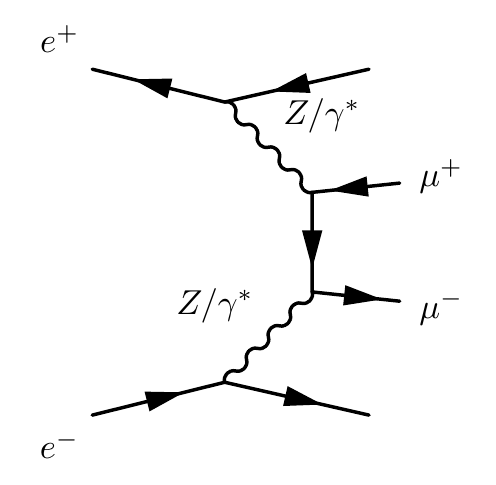}

\caption{Example of diagrams contributing significantly to the $4f$ background. Graphs such as the one on the left feature kinematics similar to the $Zh$ signal and are resonantly enhanced near the $Z$ pole. Multi-peripheral diagrams such as the one shown on the right with photons can contribute significantly in the small $-q^2$ regime ($q^\mu$ being the photon four-momentum), due to the collinear enhancement. These multi-peripheral diagrams are double counted if one naively includes both the total $4f$ background and $\gamma \gamma \to \mu^+ \mu^-$ background (in the equivalent photon approximation) without appropriate kinematic cuts.}
 \label{fig:4fbkg}
\end{figure}

Two important contributions to the $4f$ background in the signal region are shown in Fig.~\ref{fig:4fbkg}. The first corresponds to $e^+ e^- \to (Z/\gamma^*)(Z/\gamma^*)$ , with subsequent decays to four fermions including a $\mu^+ \mu^-$ pair. This process is resonantly enhanced for  $\mu^+ \mu^-$ near the $Z$ pole, and features kinematics similar to the $Zh$ signal, making it challenging to eliminate. The multi-peripheral diagrams like that shown in the right panel of Fig.~\ref{fig:4fbkg} feature a large collinear enhancement when either of the photons go close to on-shell, $-q^2 \to 0$, where $q^\mu$ is the photon 4-momentum. Although these diagrams are accounted for in principle by generating the total $e^+ e^- \to 4 f$ background all together, in practice the collinearly-enhanced low-$|q^2|$ region of the $ e^+ e^-  \to e^+ e^- + 2f$ contribution is challenging to sample in the Monte Carlo integration. We therefore break up the $4f$ event generation into two complementary regions of phase space: we generate four fermion events in \texttt{Whizard} requiring 
\beq
-(p_{e^{\pm},\, \rm initial}^{\mu} - p_{e^{\pm}, \, \rm final}^{\mu})^2 > (5 \, {\rm GeV})^2,
\eeq
where $p_{e^{\pm},\, \rm initial}^{\mu} $ is the four-momentum of the initial state electron or positron, and $p_{e^{\pm}, \, \rm final}^{\mu}$ is the four-momentum of any outgoing electron or positron in the event. This cut eliminates  $ e^+ e^-  \to e^+ e^- + 2f$ events with small $|q|^2$, but captures nearly the entire contribution from e.g.~$ZZ$ diagrams such as those on the LHS of Fig.~\ref{fig:4fbkg}. We include the effects of ISR during the event generation, as described above for the $2f$ background. We also place loose generator-level cuts requiring $m(f\overline f )>10$ GeV for all fermion-antifermion pairs. This avoids issues with soft and collinear divergences during event generation\footnote{When a $f \overline f$ pair is produced through a photon, the corresponding propagator becomes singular for vanishing invariant mass.}, but is a modest enough requirement that it should reasonably estimate the corresponding background once invariant mass cuts are imposed in our signal selection.

In the $e^+ e^- + 2f$ phase space with smaller values of $-q^2$, the multi-peripheral  contributions such as those on the RHS of Fig.~\ref{fig:4fbkg} dominate the $4f$ background. To efficiently sample these contributions, we generate $\gamma \gamma \to \mu^+ \mu^-$ events\footnote{We find that $\gamma \gamma$-initiated events with no prompt muons contribute negligibly to the signal region.} using the equivalent photon approximation in \texttt{Whizard}, discarding events in which the value of $-q^2$ for both photons exceeds (5 GeV)$^2$ to avoid double counting the phase space sampled by the $4f$ events. We also include the invariant mass cut described above for the $4f$ background, and allow finite $p_T$ to be generated for the scattered electrons in each event. 

\section{Sensitivity with and without calorimetry}
\label{sec:analysis}
With the signal and background events generated as described above, we analyze the Monte Carlo samples considering a detector with and without an ECAL or hadronic calorimeter (HCAL) system. In both cases, we assume that there is both an inner tracking and outer muon layer, separated by a layer of shielding, allowing for highly efficient muon identification and momentum determination without a significant muon fake rate, as illustrated in Fig.~\ref{fig:detector}.

\subsection{Tracker-only Analysis}\label{sec:recoil_details}

Let us first perform the analysis with tracker information only. The muons and anti-muons output by \texttt{Pythia} are taken to be reconstructed with $95\%$ efficiency provided $|\eta|<3$ and $p_T > 5$ GeV. The momentum of the muons is assumed to be determined by hits in the inner and outer tracker. To account for the finite tracker momentum resolution, we first smear all muon inverse transverse momenta, $1/p_T$, by a Gaussian centered on $(p_T/1\,{\rm GeV})^{-1}$ with width 
\beq \label{eq:track_smearing}
\sigma_{1\,{\rm GeV}/p_T} = 2\times 10^{-5} \oplus \frac{10^{-2} \, {\rm GeV}}{p_T}.
\eeq 
This resolution formula is similar to others commonly used in the $e^+ e^-$ collider literature (see e.g.~Refs.~\cite{Liu:2016zki, CEPC_Higgs}), and assumes comparable tracking technology as that envisioned for e.g.~the International Linear Detector (ILD)~\cite{ilcdetectors, CLIC_Higgs}. Elsewhere in the literature the second term in Eq.~(\ref{eq:track_smearing}) often appears with some angular dependence, as it arises from multiple scattering within the tracking material. Since it is governed by the detailed tracker geometry considered, we neglect this angular dependence in our analysis, although we have verified that our results match up well with others appearing in the literature that include a dependence on the polar angle with respect to the beam axis. Note that in smearing the muon $p_T$ by Eq.~(\ref{eq:track_smearing}), we keep the invariant mass fixed and adjust the energy accordingly. The angular variables are also held fixed, corresponding to the excellent angular resolution expected at future $e^+ e^-$ experiments. This treatment neglects inelastic interactions with the detector material, which are not expected to be significant for muons.

\begin{figure}[t!]
\begin{center}
\includegraphics[width=0.45\linewidth]{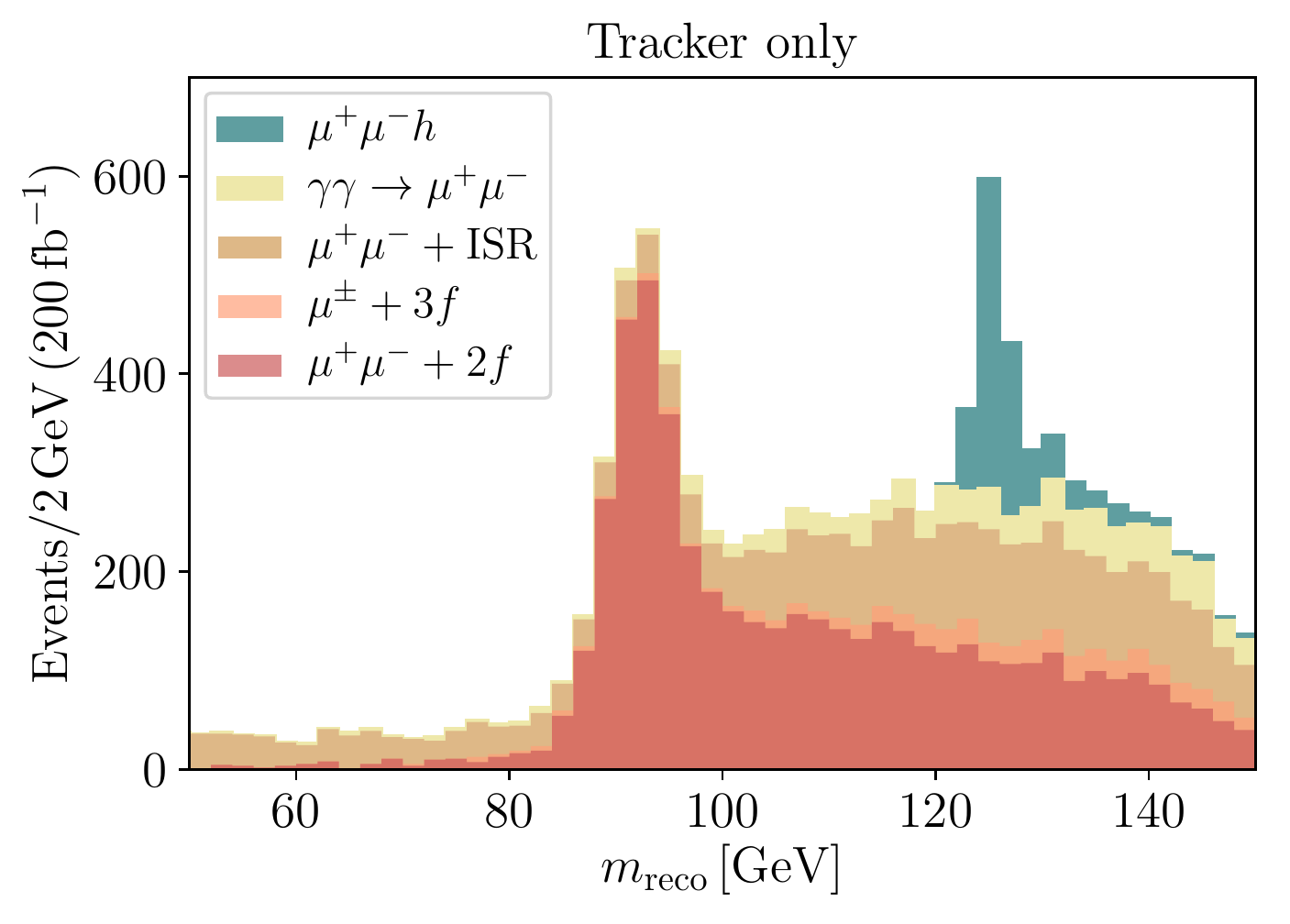} \quad \includegraphics[width=0.45\linewidth]{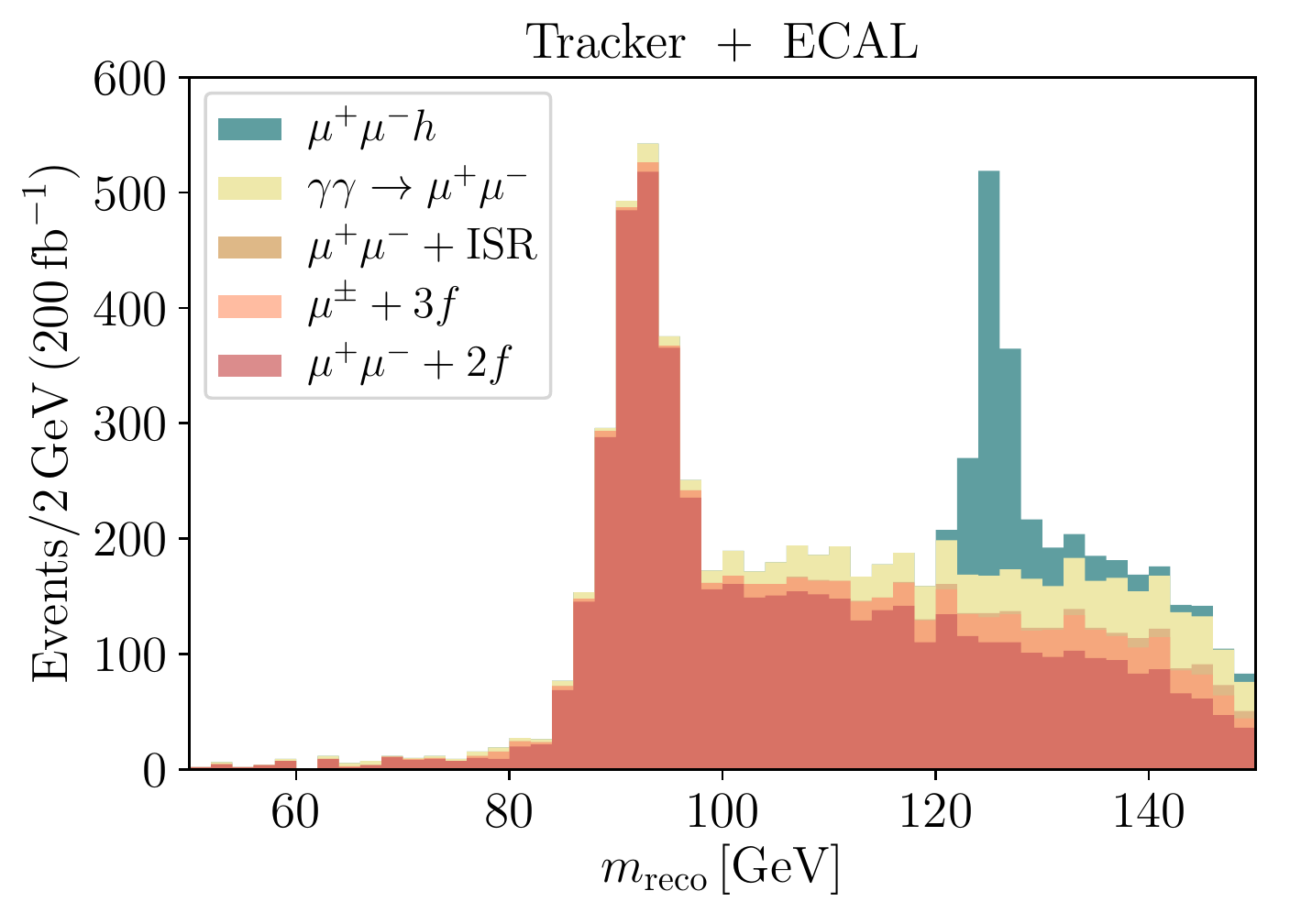} 
\caption{Distribution of $m_{\rm reco}$ for signal and background events with 200 fb$^{-1}$ after cuts (except that on $m_{\rm reco}$), with (right) and without (left) an ECAL. Calorimetry allows one to combine the FSR and muon momenta to improve the reconstruction, as well as veto on hard ISR photons, which removes almost all of the two fermion background.} 
\label{fig:mreco_dist}
\end{center}
\end{figure} 

After smearing, identified muons and antimuons are required to satisfy an isolation criterion, taken to be 
\beq
\sum_{i} \frac{ p_T(i)}{p_T(\mu^\pm)} < 0.1,
\eeq
where $i$ denotes all charged tracks in the event with $p_T > 0.1$ GeV, excluding other hard muons, within a $\Delta R < 0.2$ cone\footnote{It is more common in the lepton collider literature to use the relative lab frame angle in describing the separation between reconstructed objects, however we do not find a significant difference between this approach and utilizing the longitudinal boost-invariant angular separation $\Delta R$ in this analysis.} of the $\mu^\pm$. In this step, all charged particle momenta are also smeared by Eq.~(\ref{eq:track_smearing}) keeping the corresponding invariant masses fixed. Note that the calorimeter (energy) information is not used in this isolation criterion in either the tracker or tracker+calorimeter analyses, since we find that including calorimeter-based isolation variable does not significantly affect any of the backgrounds or signal at the level of our analysis.

For each event, we require at least one identified muon and one identified antimuon satisfying $|\eta|<3$, $p_T>5$ GeV, and Eq.~(\ref{eq:track_smearing}). If there is more than one muon and/or antimuon in an event, we choose the $\mu^+\mu^-$ pair minimizing $|m_{\mu^+\mu^-} - m_Z|$, where $m_{\mu^+\mu^-}$ is the corresponding invariant mass. We then require
\beq \label{eq:mZ_req}
m_Z-5 \, {\rm GeV} < m_{\mu^+\mu^-}  < m_Z + 5 \, {\rm GeV}
\eeq
to select events consistent with a $Z\to \mu^+ \mu^-$ decay. To isolate events consistent with the decay of a Higgs, we compute the recoil mass
\beq \label{eq:reco_cut}
m_{\rm reco} \equiv \sqrt{s +m_{\mu^+\mu^-}^2 - 2E_{\mu^+\mu^-}\sqrt{s}}
\eeq 
where $E_{\mu^+\mu^-}$ is the sum of the selected muon and antimuon energies. Genuine $\mu^+ \mu^- h$ events will have $m_{\rm reco} \approx m_h$, and so we require
\beq
120\,{\rm GeV} < m_{\rm reco}<140\,{\rm GeV}.
\eeq
Since most signal events will feature moderately transverse muons from recoiling against a Higgs, we impose additional transversality cuts on the $Z$ candidate, requiring
\beq
 p_T(\mu^+, \mu^-) >30 \, {\rm GeV}, \quad p_L(\mu^+, \mu^-) <60 \, {\rm GeV}.
\eeq
This significantly reduces many of the backgrounds, especially $2f + {\rm ISR}$. Also, since the $Z$ candidate in genuine $Zh$ events tends to be less boosted than for the backgrounds, we place a requirement on the acollinearity of the muons,
\beq \label{eq:acol}
\cos^{-1} \left(\frac{\mathbf{p}_{\mu^-} \cdot \mathbf{p}_{\mu^+}}{\left| \mathbf{p}_{\mu^-} \right| \left|\mathbf{p}_{\mu^+}\right|} \right) > 100^\circ.
\eeq
For illustration, Fig.~\ref{fig:mreco_dist} shows the distribution of $m_{\rm reco}$ for the various background and signal contributions after all cuts (except that on $m_{\rm reco}$). With only tracker information, the Higgs signal peak is distinct.

The number of expected signal and background events expected with 200 fb$^{-1}$ integrated luminosity after all cuts listed above as well as requiring 120 GeV$<m_{\rm reco}<130$ GeV are shown in the first row of Table~\ref{tab:recoil}. The background is dominated by $4f$ events ($\approx 49\%$), followed by the $\mu^+\mu^-+{\rm ISR}$ contribution ($\approx 38\%$); the $\gamma\gamma$--induced background contributes about $13\%$. We again emphasize that the requirements imposed so far only make use of tracker information. 

For binned signal and background samples, and accounting only for statistical uncertainties, the expected sensitivity to $\sigma_{Zh}$ is approximately given by
\beq
\frac{\delta \sigma_{Zh}}{\sigma_{Zh}} \simeq \left( \, \sum_{i\in {\rm bins}} \frac{S_i^2}{S_i + B_i} \,\right)^{-1/2}
\eeq
With the tracker-only requirements reflected in the first row of Table~\ref{tab:recoil}, and considering the samples with 2 GeV bins in $m_{\rm reco}$, we find that a precision of about 5.5\% (1.1\%) can be achieved in $\sigma_{Zh}$ with 200 fb$^{-1}$ (5 ab$^{-1}$).

The cuts reflected in Eqs.~(\ref{eq:mZ_req})--(\ref{eq:acol}) are similar to those appearing in several past analyses in the context of the ILC~\cite{ILC_Higgs,Yan:2016xyx}, the FCC-ee~\cite{tlepwg, LEP3}, CLIC~\cite{CLIC_Higgs}, and the CEPC~\cite{Chen:2016zpw, CEPC_Higgs}. Additional requirements can be imposed on the reconstructed $Z$ system that could potentially further reduce backgrounds  (e.g. cuts on the acoplanarity of the $Z$ or angular separation of the muons). However, for the $Z\to \mu^+ \mu^-$ channel, we do not find these additional cuts to provide significant gains in sensitivity to $\sigma_{Zh}$ in our cut-based analysis. Also, more sophisticated multivariate methods have been used in the literature to further reduce some of the remaining backgrounds~\cite{Chen:2016zpw, CLIC_Higgs}. However, since these methods typically rely solely on information about the $\mu^+\mu^-$ momenta, obtained from the tracker and muon system, we do not expect these improvements to significantly impact our comparison between the tracker and tracker+calorimeter performance in the recoil measurement.

\begin{table}
\centering
\begin{tabular}[!t]{c || c  c   || c c}

Process: & $Zh$ signal & Background & $\frac{\delta \sigma_{Zh}}{\sigma_{Zh}} $ (200 fb$^{-1}$) & $\frac{\delta \sigma_{Zh}}{\sigma_{Zh}} $ (5 ab$^{-1}$) \\

\hline
\hline

Tracker only: &$634$  & $1372$   & 5.9\% & 1.2\%  \\

Tracker + ECAL: & $704$ & $874 $  &4.9\% & 1.0\% \\

\end{tabular}

\caption{Approximate number of signal and background events expected with 200 fb$^{-1}$ of integrated luminosity in an analysis with and without calorimetry, as well as the projected sensitivities to the $Zh$ cross-section assuming 200 fb$^{-1}$ and 5 ab$^{-1}$. The signal and background numbers above include an additional mass window requirement 120 GeV $<m_{\rm reco}<$ 130 GeV. The ECAL allows for a veto on hard ISR photons which significantly reduces the $2f$ background, as well as the recovery of final state photons (produced as final state radiation through \texttt{Pythia} in our analysis) which improves the signal reconstruction and background rejection.  \label{tab:recoil}}
\end{table}

We can also estimate the sensitivity of tracking detectors to the Higgs mass. To do so, we characterize the precision achievable in determining the Higgs mass by the expected statistical precision achievable in the measured mean value of $m_{\rm reco}$:
\beq\label{eq:mean_mass}
\langle m_h \rangle = \frac{\sum_i m_{\rm reco, i} s_i}{\sum_j s_i}.
\eeq
Here $s_i$ and $m_{\rm reco,i}$ are the number of signal events and value of $m_{\rm reco}$ corresponding to the $i$th bin, and the sum is over bins in a window around  $m_h=125$ GeV, the mass of the Higgs used as input for the Monte Carlo event generation. Near $m_h$, the signal in the $m_{\rm reco}$ distribution is expected to be approximately Gaussian, with a non-Gaussian tail extending to higher masses due to initial and final state radiation (see Fig.~\ref{fig:mreco_dist}). We therefore restrict the sum above to be near the Gaussian core. The statistical precision in $\langle m_h\rangle$, denoted as $\delta m_h$, can be expressed as
\beq \label{eq:m_precision}
\left(\delta m_h  \right)^2 = \frac{\sum_i \left(s_i + b_i \right)\left(m_{\rm reco,i} - \langle m_h \rangle\right)^2}{\left(\sum_j s_i\right)^2}
\eeq 
where $b_i$ is the number of background events in the $i$th bin, and the sum extends over the same range as in Eq.~(\ref{eq:mean_mass}). Considering the signal in a 3 GeV window centered on $m_h = 125$ GeV, we find that the value of $\delta m_h $ for the analysis with a tracking detector and without calorimetry is expected to be
\beq\label{eq:m_tracking}
\delta m_h \simeq 56 \, {\rm MeV} \,\, ({\rm 200 \, fb}^{-1}).
\eeq 
This result does not depend significantly on the bin size. As we will show below, the result of Eq.~(\ref{eq:m_tracking}) is comparable to the precision achievable including calorimeters.

\subsection{Adding Calorimeters}\label{sec:muons_w_calo}

So far, the entire analysis described above can apply to a detector with or without an ECAL/HCAL. Since the recoil measurement is specifically designed to be independent of the Higgs decay mode, the cuts in Eqs.~(\ref{eq:mZ_req})--(\ref{eq:acol}) depend only on the muon four-momenta, which can be measured to excellent precision with the tracker and muon system. How can this analysis benefit from calorimetry?

The most important handle provided by an ECAL in this analysis is sensitivity to photons. As a result, photons radiated off of the $Z$ candidate muons can be accounted for to improve the efficiency for reconstructing the $Z$. To investigate this effect, we re-analyze our Monte Carlo events combining final state muon and photon four-momenta for photons within $\Delta R <0.2$ of the $\mu^\pm$ reconstructing the $Z$, and subsequently removing the corresponding photons from the event record. Photon momenta would be measured by the ECAL with finite resolution, and so following e.g.~\cite{Liu:2016zki, CEPC_Higgs}, we smear photon energies by a Gaussian centered on $E_\gamma/1{\rm GeV}$ with width
\beq \label{eq:ECAL_res}
\sigma_E^{\rm ECAL} = 0.16\sqrt{\frac{E}{\rm 1\, GeV}} \oplus 0.01 \frac{E}{\rm 1\, GeV},
\eeq
keeping the momentum direction and vanishing invariant mass fixed.  
 The signal peak is somewhat increased due to the improved reconstruction efficiency, as can be seen in the right panel of Fig.~\ref{fig:mreco_dist} which displays the $m_{\rm reco}$ distributions incorporating calorimetry, however the resulting improvement is not very significant. Note that our analysis does not include the effects of energy loss from  interactions of muons with the material (occurring through e.g.~ionization, atomic excitations, and bremsstrahlung), however these effects are expected to be small for muons (i.e.~they are minimum ionizing particles). Accounting for FSR and bremsstrahlung can be important for electrons in the $e^+e^-\to Zh\to e^+e^-+X$ analysis, however, as discussed briefly below in Sec.~\ref{sec:electrons}. 

Sensitivity to photons also allows for the reduction of backgrounds in the muon analysis. As mentioned above, the large $2f+{\rm ISR}$ background can only contribute to the signal region provided at least one photon with substantial transverse momentum. In a detector with an ECAL, one might veto on sufficiently hard, transverse photons to mitigate this background. This approach was taken e.g.~in the TLEP/FCC analysis~\cite{tlepwg, LEP3}. To investigate this effect, we analyze photons generated in events passing the signal criteria above and exclude events from our tracker+ECAL analysis if they have any photons with
\beq
p_T(\gamma) > 30 \,{\rm GeV}, \quad \left|\eta\right|<5.
\eeq
Note that recovered FSR photons are not considered in this veto.

The effect of these improvements on the $e^+e^-\to Zh\to\mu^+\mu^-+X$ analysis is reflected in the second row of Table~\ref{tab:recoil}. The signal efficiency is increased, owing to the improved reconstruction of the $Z$ by combining the FSR and muon momenta. The total background is reduced by about 40\%, due primarily to the reduction in the $2f+{\rm ISR}$ contribution, which becomes negligible in the signal region (see also the right panel of Fig.~\ref{fig:mreco_dist}). The photon veto reduces the signal by $< 1\%$. We find that a sensitivity of $\delta_{Zh}/\sigma_{Zh} \simeq 4.9\%$ can be achieved with these improvements made possible by calorimetry, assuming 200 fb$^{-1}$. This corresponds to a $\sim 17\%$ improvement over the tracker-only projection. Sensitivity to the $m_h$ can also be characterized by Eq.~(\ref{eq:m_precision}) and compared to the results of our tracker-only analysis. We find that with the improvements discussed above, a precision of
\beq
\delta m_h \simeq 47 \, {\rm MeV} \,\, ({\rm 200 \, fb}^{-1}).
\eeq
can be achieved, considering the signal in the same window as in Eq.~(\ref{eq:m_tracking}). This corresponds to a modest $16\%$ improvement over the analysis with a tracking detector.

Another potential benefit of an ECAL and HCAL in the muon recoil analysis is the ability to measure the total visible energy and momentum in an event. Some previous studies impose requirements on the polar angle of the missing momentum, $\theta_{\rm miss}$ (see e.g.~\cite{Yan:2016xyx}). By placing a lower bound on $\theta_{\rm miss}$, events with hard ISR photons outside of the detector acceptance can be rejected. We have investigated this effect in our analysis, and do not find a significant improvement over the results without missing momentum cuts once the photon veto and final state photon recovery are taken into account. 

Apart from photon identification and measuring missing momentum, we are not aware of any additional information relying on calorimetry that would significantly improve $\delta \sigma_{Zh}/\sigma_{Zh}$ in the muon channel. Of course, in other channels this conclusion does not necessarily hold. We are also not aware of any additional backgrounds that would become relevant without calorimetry (provided high muon ID and fake rejection efficiencies can be achieved with the tracker alone). 

\subsection{Alternative methods for photon identification} \label{sec:alt}

Even without detector calorimetry, it may be possible to infer the presence of photons in an event and effectively eliminate events with hard photons from the signal region, mimicking a photon veto. For example, one could include a layer of lead a few radiation lengths thick outside the inner tracker, and enclosed by an additional tracking layer within the solenoid. Photons impinging on the lead layer would convert into e.g.~$e^+ e^-$ pairs, which would be ejected into the second tracking layer and be detected. If converted electrons and positrons were observed in the second tracking layer without associated tracks in the inner detector, one would infer the presence of either a photon or a jet containing only neutral hadrons. The fraction of hadronic jets without any associated charged tracks is expected to be small. For the $\mu^+ \mu^- h$ signal, we have verified that for $p_T(j)\gtrsim 5$ GeV, the fraction of trackless jets is negligible, while for lower $p_T$ the fraction can be at the percent level. Thus, converted $e^+ e^-$ pairs with significant $p_T$ observed in the second tracking layer without associated tracks in the inner tracker likely indicates the presence of a hard photon, and the corresponding event can be vetoed. 

A more detailed analysis of this strategy would involve a simulation of photon interactions with lead, and is beyond the scope of this work. Nevertheless, it appears possible to obtain some information about the presence of photons using a variation of this approach. If so, the sensitivity of the tracking detectors to $\delta\sigma_{Zh}$ can approach that suggested in the second row of Table~\ref{tab:recoil} and the RHS of Fig.~\ref{fig:mreco_dist} even without calorimetry.

\subsection{Summary of $e^+e^-\to Zh\to\mu^+\mu^-+X$}

With tracker information alone, a future $e^+ e^-$ collider with 200 fb$^{-1}$ of integrated luminosity could probe $\sigma_{Zh}$ to a precision of around 5.9\% (1.1\%) given 200 fb$^{-1}$ (5 ab$^{-1}$) of integrated luminosity, compared to around 4.9\% (1\%) with FSR recovery and a photon veto allowed by including calorimetry. The Higgs mass can also be measured to comparable precision: $\delta m_h \simeq 56$ MeV (47 MeV) without (with) calorimetry, and assuming 200 fb$^{-1}$. Sensitivities similar to that of our tracking+calorimetry analyses might be achieved without an ECAL by adding lead shielding between the inner detector and an extra tracking layer to veto photon conversions. 
 Thus, we conclude that a stage of tracker-only operation at a future $e^+ e^-$ collider could come close to the sensitivity afforded by a full detector and already significantly improve over what can be done at the LHC, at least in the $\mu^+ \mu^-$ channel.
 
\section{Comments on $e^+e^-\to Zh\to e^+ e^- +X$}
\label{sec:electrons}
We have shown that relatively high precision can be obtained for the $Zh$ recoil measurement in the muon channel without calorimeters. This is because the most important selection criteria, cuts, and observables depend only on the kinematics of the muons, which can be accurately identified and measured with the tracking system. Here, we comment briefly on the prospects for performing an electron recoil measurement without calorimeters, deferring a full analysis of the electron recoil measurement to future work.

Consider the process
\beq
e^+ e^- \to Z h, \qquad Z \to e^+ e^-, \, h\to {\rm anything}.
\eeq
The situation is more complicated for electrons than for muons without detector calorimetry. One challenge is that electrons and charged hadrons must be distinguished with tracker information only. How well might the tracker identify the difference between e.g.~an $e^-$ and a $\pi^-$?

\begin{figure}[t!]
\begin{center}
\includegraphics[width=0.55\linewidth]{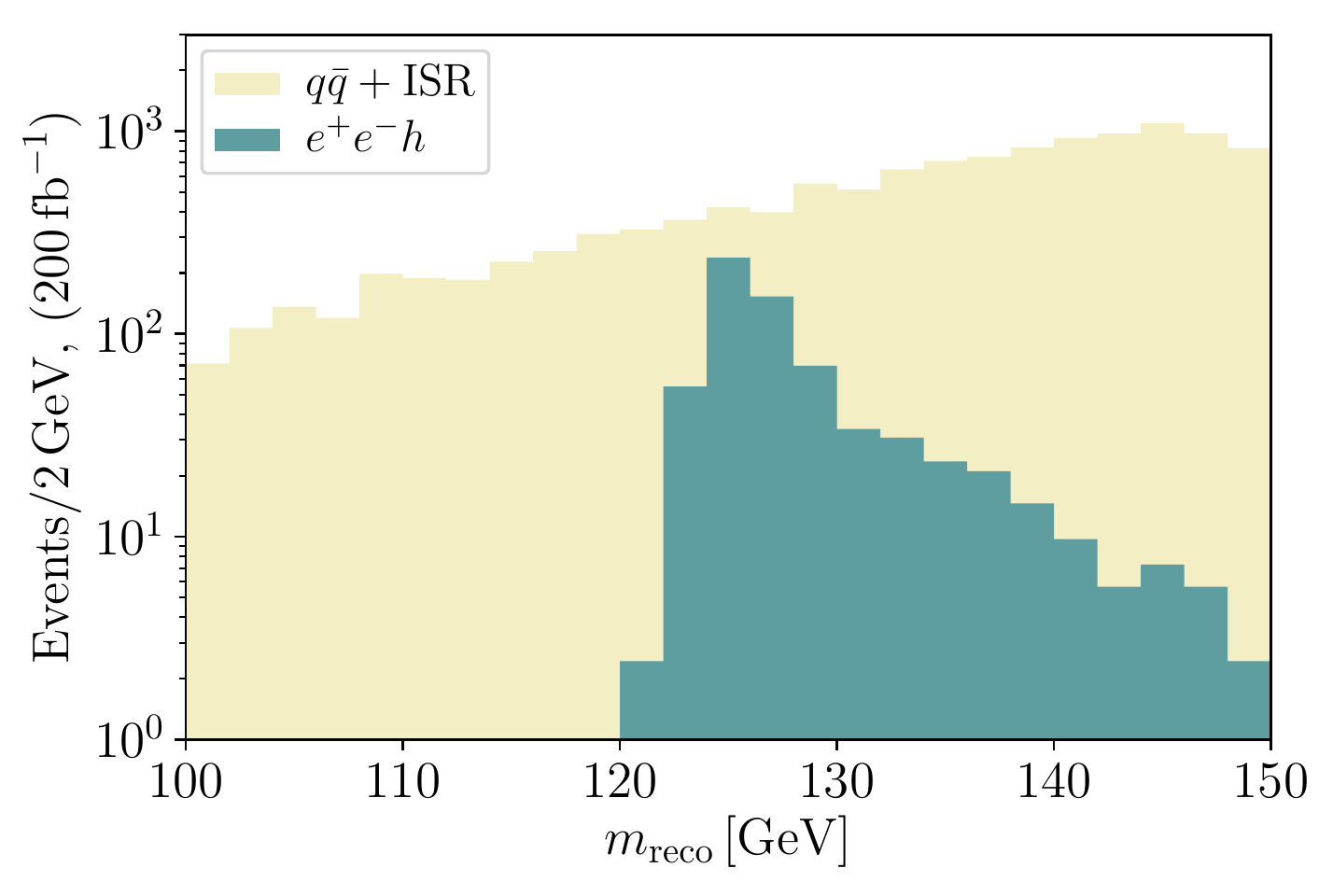} 
\caption{Comparison of the $m_{\rm reco}$ distribution for the $Z(\to e^+ e^-)h$ signal and $q\bar{q}+{\rm ISR}$ background with tracker information only. At this level, no track isolation requirements have been imposed on the electron candidates, resulting in a large $q \bar{q}$ contribution where charged hadrons fake electrons. This background can be reduced by the track isolation requirements discussed in the text.} 
\label{fig:mreco_electrons}
\end{center}
\end{figure} 

As a first step in addressing this question, we simulate $e^+ e^- \to e^+ e^- h$ events, as well as $e^+ e^- \to q \bar{q} + {\rm ISR}$. The $q \bar{q}$ background is large, and could potentially bury the $Zh$ recoil signal if electrons cannot be adequately distinguished from jets.  We perform an analysis similar to that described in Sec.~\ref{sec:recoil} for muons. First, we define electron candidates\footnote{In this analysis, we do not distinguish between the charges of the corresponding tracks.} simply as tracks with $p_T > 10$ GeV, $|\eta| < 3$, and with no corresponding hit in the muon system. At this level, all track $p_T$s output by \texttt{Pythia} are smeared by Eq.~(\ref{eq:track_smearing}). Note that the effects of inelastic interactions of the charged particles with the tracking material (e.g.~bremsstrahlung) are neglected in what follows, but would be important to address in a full analysis. After smearing, we proceed as before, requiring at least two electron candidates per event, then reconstructing the $Z$ out of the candidates that minimize $|m_{e^+ e^-}-m_Z|$, and requiring 
\beq
p_T(e^+, e^-) >30 \, {\rm GeV}, \quad p_L(e^+, e^-) <60 \, {\rm GeV}, \quad \cos^{-1} \left(\frac{\mathbf{p}_{e^-} \cdot \mathbf{p}_{e^+}}{\left| \mathbf{p}_{e^-} \right| \left|\mathbf{p}_{e^+}\right|} \right) > 100^\circ,
\eeq
in analogy with our selection criteria in the muon case. We compute $m_{\rm reco}$ from Eq.~(\ref{eq:reco_cut}) with $m_{\mu^+ \mu^-}$, $E_{\mu^+ \mu^-} \to m_{e^+ e^-}$, $E_{e^+ e^-}$ and require 120 GeV$<m_{\rm reco}<130$ GeV. The corresponding $m_{\rm reco}$ distributions for the $Zh$ signal and $q \bar{q}$ background are shown in Fig.~\ref{fig:mreco_electrons} after all cuts except those on $m_{\rm reco}$. The signal peak is still evident, but lies beneath the $q \bar{q}$ background at this stage. At this level, no track isolation requirements have been imposed on the electron candidates.

\begin{figure}[t!]
\begin{center}
\includegraphics[width=0.45\linewidth]{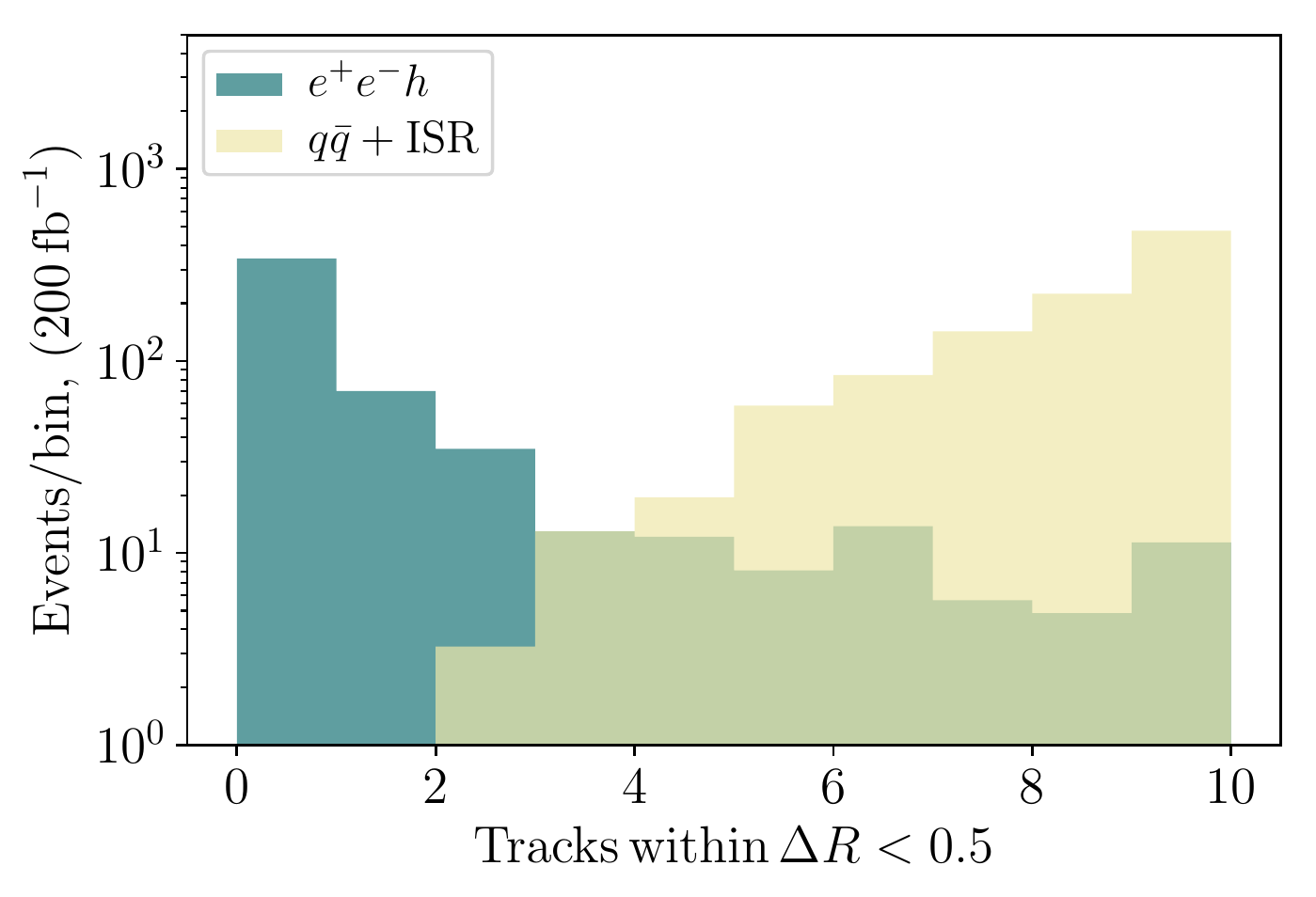}\,\includegraphics[width=0.45\linewidth]{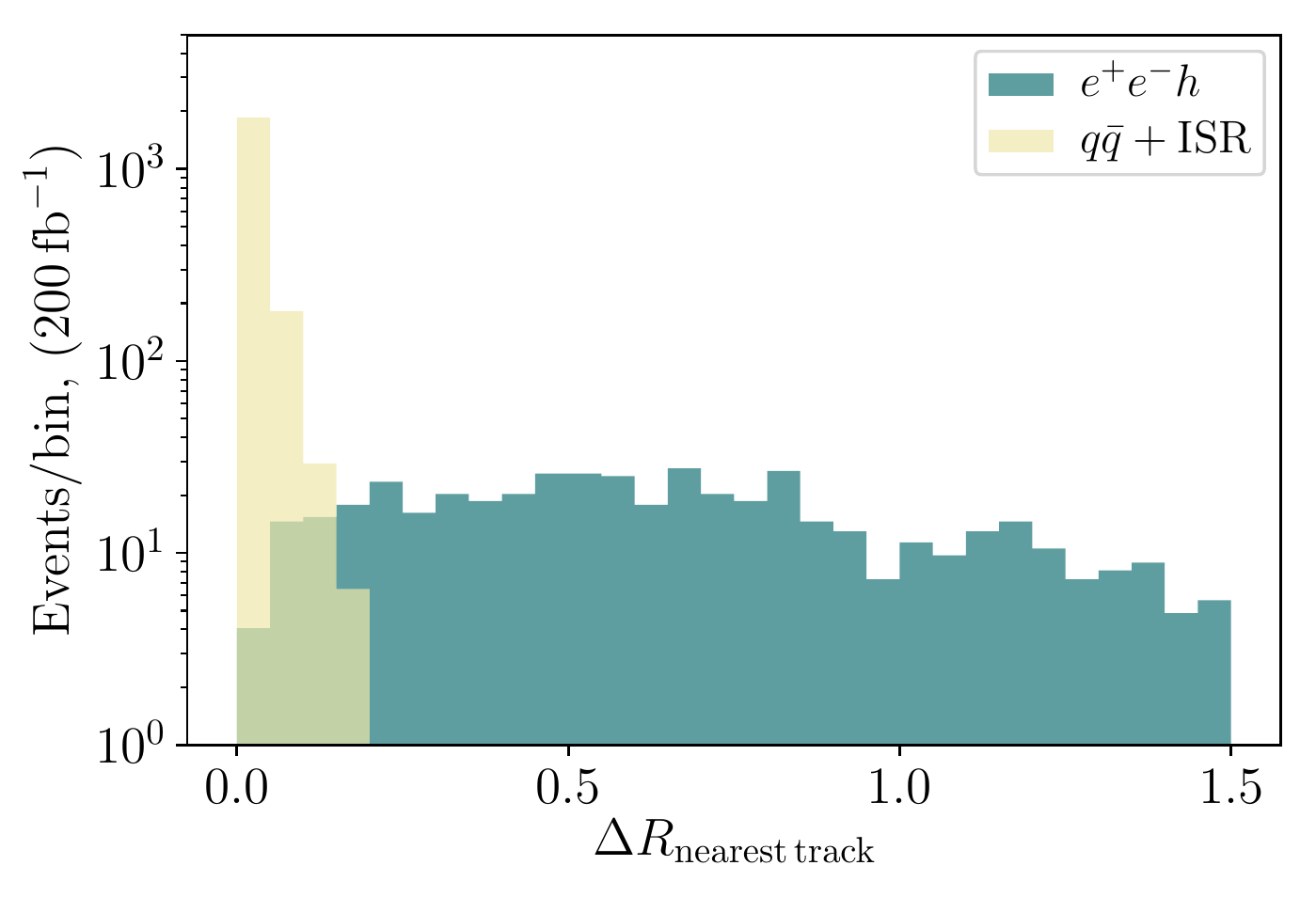} 
\caption{Track isolation information for the $Z(\to e^+ e^-)h$ signal and $q\bar{q}+{\rm ISR}$ background. On the left we show a histogram of the total number of charged tracks with $p_T > 0.5$ GeV within a cone of $\Delta R = 0.5$ around either electron candidate in each event passing all other cuts (including the $m_{\rm reco}$ window requirement). All $e^+ e^-$ candidates in our generated event sample of $q \bar{q}$ events passing cuts have several charged tracks within $\Delta R < 0.5$, reflecting the effects of QCD radiation that is not present for genuine electrons. On the right we show the distance between the least isolated electron candidate and the nearest charged track with $p_T>0.5$ GeV. Once again, genuine electrons tend to be more isolated, while the generated $q\bar{q}$ background always features a charged track within $\Delta R\lesssim 0.2$ of one of the electron candidates.} 
\label{fig:electrons}
\end{center}
\end{figure} 

For the $q \bar{q}$ background, QCD radiation typically results in electron candidates that are constituents of jets featuring several nearby charged tracks. In contrast, genuine electrons tend to propagate in isolation from other hard charged particles. We can therefore distinguish between genuine electrons and charged hadrons by considering tracker-based isolation criteria. To characterize the isolation, we consider observables that depend on the longitudinal boost-invariant angular distance $\Delta R$ instead of the more conventional relative lab frame angle, since $q\bar{q}$ events contributing to the signal region can feature a substantial longitudinal boost. On the left hand side of Fig.~\ref{fig:electrons}, we show the total number of charged tracks with $p_T > 0.5$ GeV within a cone of $\Delta R = 0.5$ around either electron candidate in each event passing all other cuts (including the $m_{\rm reco}$ window requirement). All generated $q \bar{q}$ events passing cuts have at least 2 additional charged tracks within $\Delta R < 0.5$ of the electron candidates, reflecting the effects of QCD radiation that is not present for genuine electrons. On the right hand side of Fig.~\ref{fig:electrons}, we show the minimum distance between either electron candidate and a charged track with $p_T>0.5$ GeV, again after all cuts. The simulated $q\bar{q}$ background always features a charged track within $\Delta R\lesssim 0.2$ of one of the electron candidates, while $\approx 90\%$ of the signal features more isolated electrons. Therefore, requiring e.g.~$\Delta R_{\rm nearest \, track} > 0.2$ would significantly reduce the $q \bar q + {\rm ISR}$ background, while only marginally affecting the signal. We expect similar conclusions for the other multi-jet backgrounds. Therefore it may indeed be possible to obtain good sensitivity to $\delta \sigma_{Zh}$ in the electron channel without detector calorimetry, although a more detailed analysis is required to draw firm conclusions. 

There are other challenges associated with a tracker-only analysis of $e^+e^-\to Zh\to e^+ e^- +X$ that should be kept in mind. For one, the backgrounds are larger than for the muon case, since there are many more topologies with an $e^+ e^-$ pair in the final state (although this is true for the analysis with calorimetry as well). Also, electrons tend to produce more final state radiation and bremsstrahlung photons than muons, decreasing the $Z$ reconstruction efficiency and degrading the Higgs signal peak if the radiated photons are not accounted for~\cite{Yan:2016xyx, CLIC_Higgs}. Properly analyzing the latter effect requires accounting for inelastic interactions of electrons in the material, which we have not included. We hope to address these issues and provide a more definitive analysis of  $e^+e^-\to Zh\to e^+ e^- +X$ with and without calorimetry in a forthcoming publication.

\section{Light Scalar Searches}\label{sec:light_scalar}

\begin{figure}[t!]
\begin{center}
\includegraphics[width=0.45\linewidth]{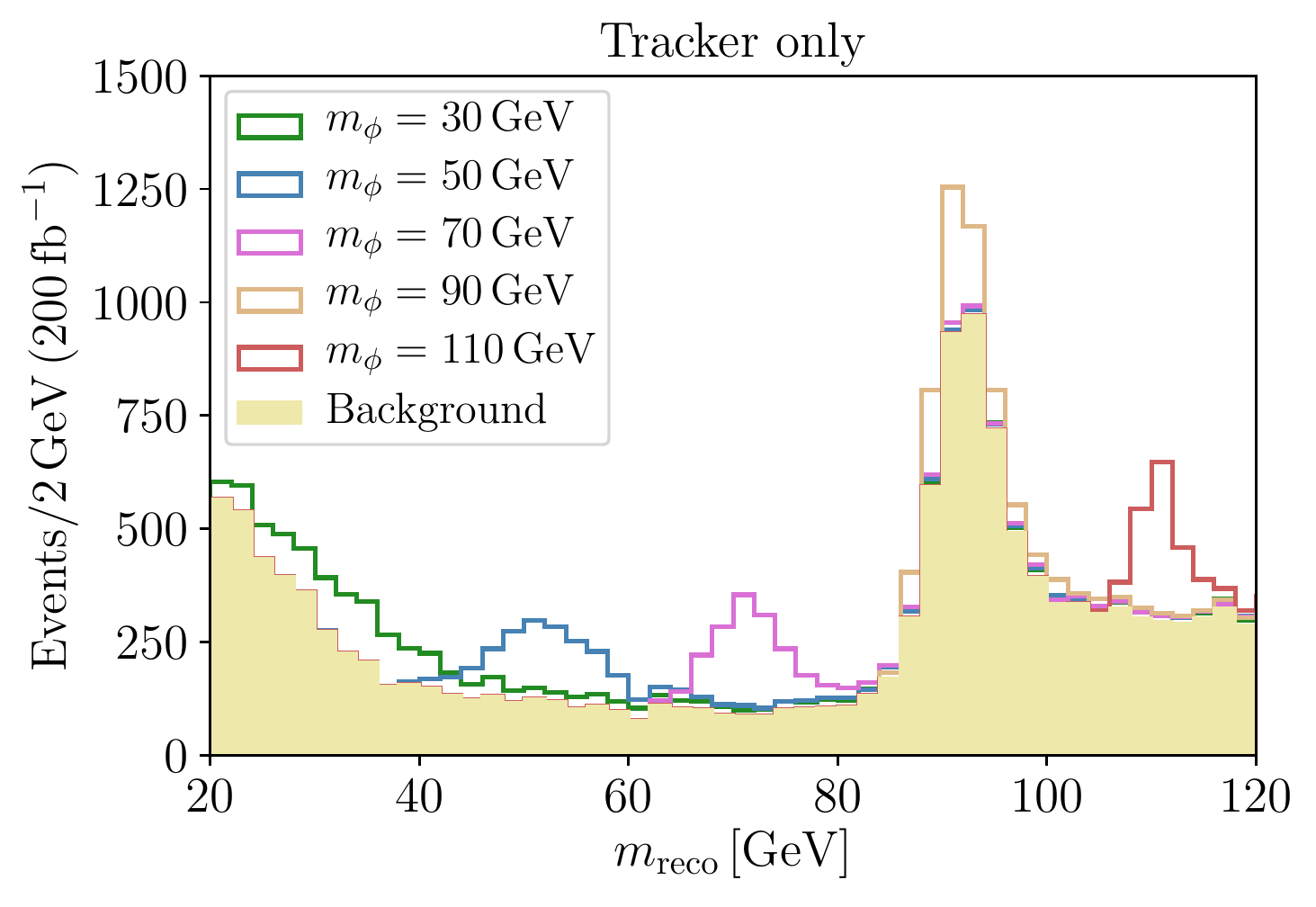} \quad \includegraphics[width=0.45\linewidth]{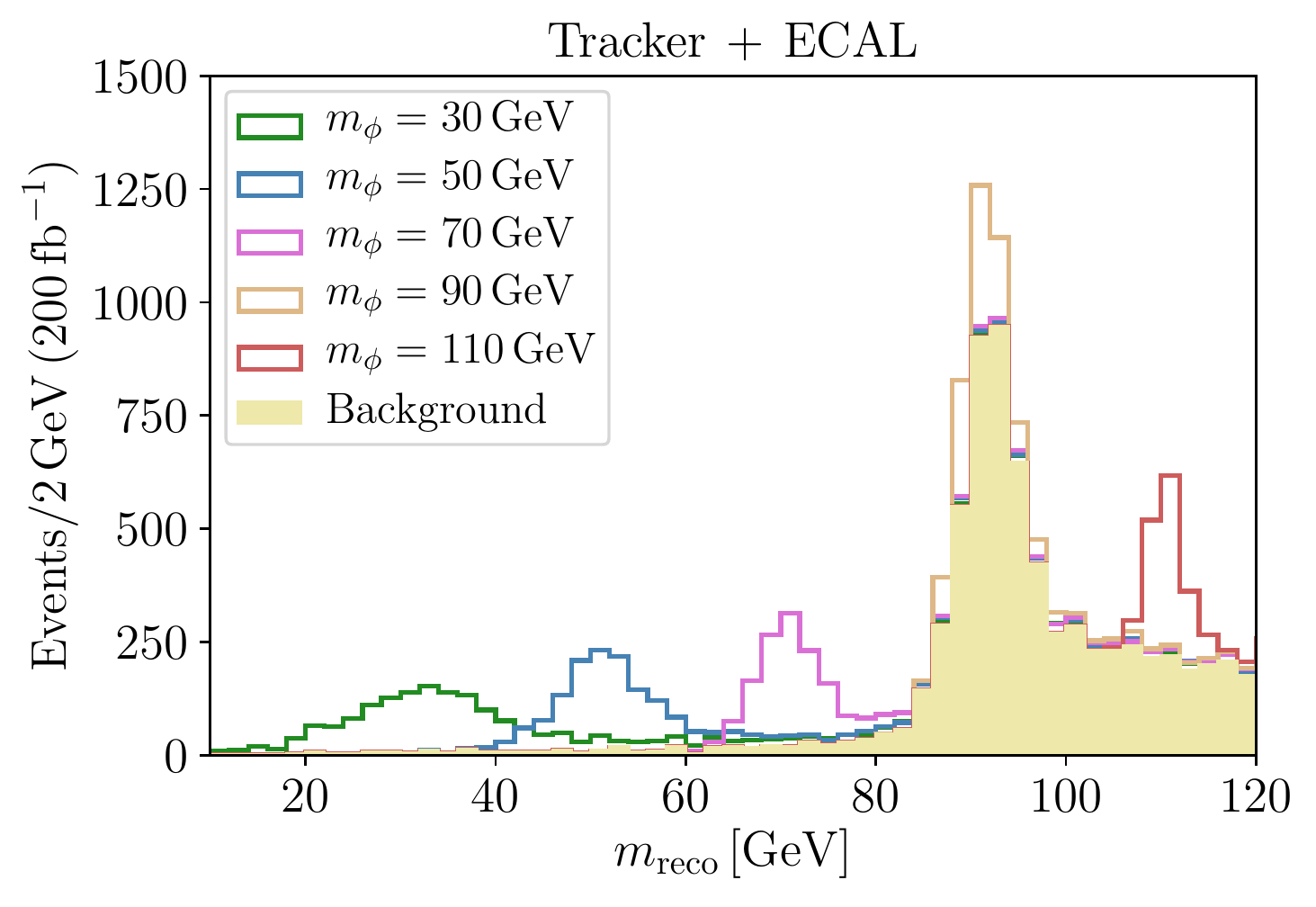} 
\caption{$m_{\rm reco}$ distributions for a light scalar recoil search in the $Z\to\mu^+\mu^-$ channel with 200 fb$^{-1}$ after cuts. Results are shown for several scalar masses with $\kappa = 1$. The plot on the right (left) shows the distributions with (without) an ECAL.} 
\label{fig:mphi_dist}
\end{center}
\end{figure} 

It is also worthwhile to consider the detector capabilities required in searching for light scalars beyond the SM Higgs at future $e^+ e^-$ colliders. A decay mode--independent search can be performed using the recoil method in analogy with the $Zh$ measurement (see e.g.~\cite{Wang:2018fcw} for a similar study for the ILC). This strategy was used at LEP by the OPAL collaboration to set limits on light scalars coupled to the $Z$~\cite{OPAL}. Here, we compare the sensitivities achievable in the $Z\to \mu^+\mu^-$ channel with and without calorimetry. Note that in concrete models involving new light scalars, other searches targeting specific decay modes of the $Z$ and Higgs can be more sensitive than the decay mode--independent search (see e.g.~\cite{Barate:2003sz,Chang:2017ynj, Chang:2018pjp}).

For our analysis, we consider a scalar $\phi$ with mass $m_{\phi}<m_h$ that couples to the $Z$ with strength reduced by $\kappa$ with respect to the SM Higgs, so that $\sigma_{Zh}$ is reduced by $\kappa^2$ relative to that expected for a SM Higgs of the same mass. For models in which $\phi$ inherits its couplings to the SM by mixing with the Higgs, $\kappa^2 = \sin^2\theta$, where $\theta$ is the $\phi-h$ mixing angle. For the recoil analysis we can remain agnostic about the decay modes of $\phi$, although in our Monte Carlo we allow it to decay as a SM-like Higgs of the same mass for simplicity. As for the SM case, we generate signal events in \texttt{Whizard 2} incorporating the effects of ISR, and use \texttt{Pythia 6} for showering and hadronization. To estimate sensitivities, we make use the background sample from the $Zh$ analysis. 

\begin{figure}[t!]
\begin{center}
\includegraphics[width=0.55\linewidth]{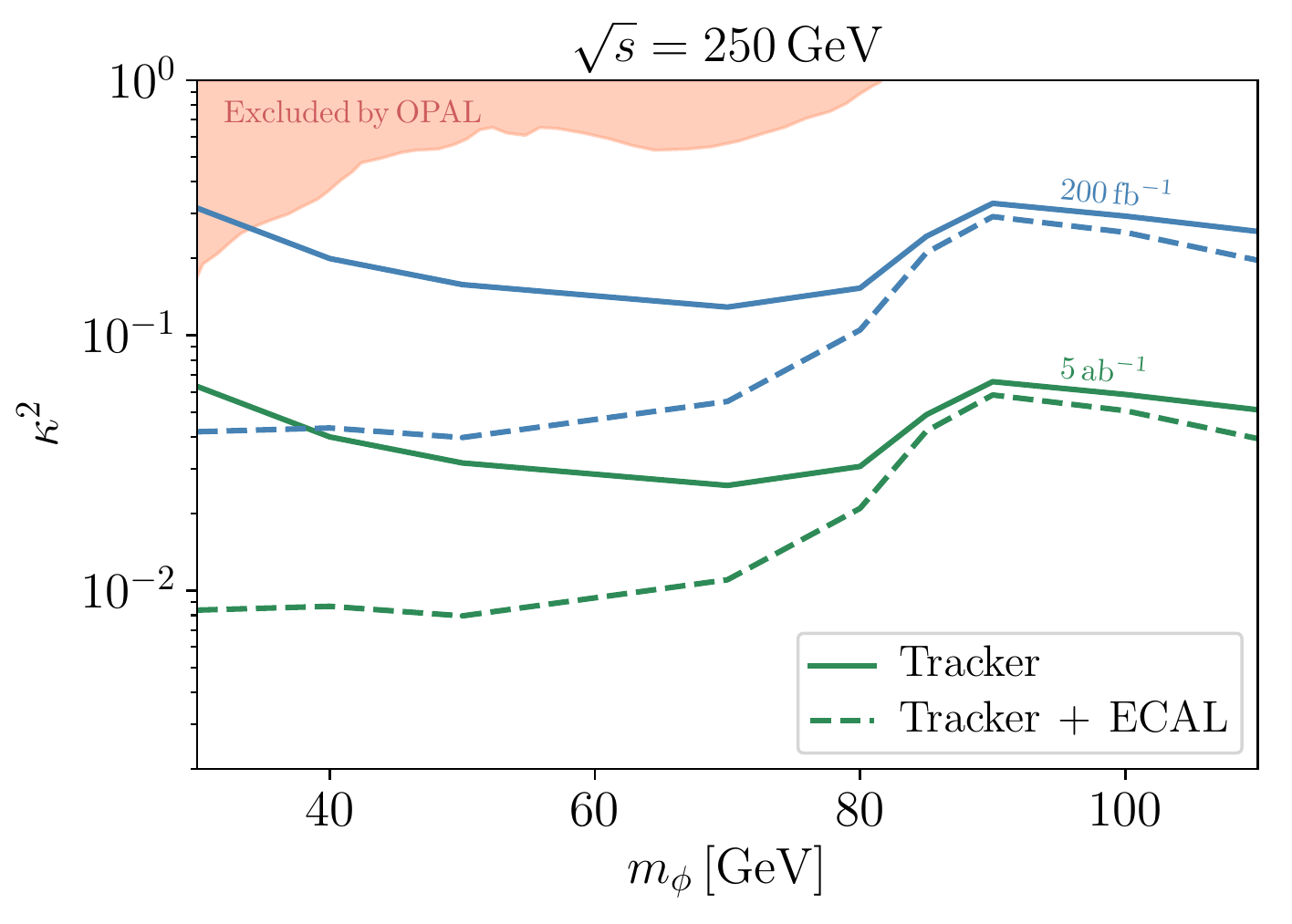}  
\caption{Discovery reach for new light scalars in a $Z(\rightarrow \mu^+ \mu^-) \phi$ recoil search at a future Higgs factory with 200 fb$^{-1}$ and 5 ab$^{-1}$ of integrated luminosity. The dashed (solid) curves show the sensitivity to the effective $\phi-Z$ coupling with (without) calorimetry. In the tracker-only case, the reach is degraded at low masses due to the significant $2\mu + {\rm ISR}$ background, which can be eliminated by a photon veto or the alternative photon ID method described in Sec.~\ref{sec:alt}. Also shown for reference are the analogous decay mode--independent bounds from OPAL~\cite{OPAL}.} 
\label{fig:scalar}
\end{center}
\end{figure} 

Our selection criteria mimic those of the $Z(\to \mu^+\mu^-)h$ measurement, only now we require $m_{\rm reco} < 120$ GeV and loosen the acollinearity cut, as lighter scalars are produced with larger characteristic boost:
\beq \label{eq:acol_2}
\cos^{-1} \left(\frac{\mathbf{p}_{\mu^-} \cdot \mathbf{p}_{\mu^+}}{\left| \mathbf{p}_{\mu^-} \right| \left|\mathbf{p}_{\mu^+}\right|} \right) > 40^\circ.
\eeq
To estimate the sensitivities to $\kappa$, we do a simple comparison of the total signal ($S$) and background ($B$) in a 15 GeV window centered around $m_{\phi}$ for each mass, defining the discovery reach in $\kappa$ as those values predicting $S/\sqrt{B}\geq 5$. This analysis neglects systematic uncertainties, but provides a reasonable comparison of the sensitivities achievable with and without calorimetry.

The recoil mass distributions for the SM background and signal for various values of $m_{\phi}$ and $\kappa = 1$ are shown in Fig.~\ref{fig:mphi_dist} after applying all cuts except the $m_{\rm reco}$ requirement. The plot on the right (left) shows the distributions with (without) an ECAL. Note that we only consider scalar masses down to 30 GeV, as going below this requires looser cuts than those assumed above. The signal peak is wider at lower masses, corresponding to the decreased $p_T$ resolution for more boosted muons, reflected in Eq.~(\ref{eq:track_smearing}). In the tracker-only case, the background is significantly larger below $m_W$, due to the $2\mu + {\rm ISR}$ contribution. This degrades the sensitivity to scalars with $m_\phi \lesssim m_W$, as shown in Fig.~\ref{fig:scalar}, which illustrates the corresponding discovery potential in the $Z(\rightarrow \mu^+ \mu^-) \phi$ recoil search with and without calorimetry. Also shown for reference are the current limits from the analogous decay mode--independent search at OPAL~\cite{OPAL}. Although the sensitivity to light scalars below $\sim m_W$ is reduced without a photon veto, the alternative photon ID method discussed in Sec.~\ref{sec:alt} could be used to improve the tracker-only reach so that it is comparable to that afforded by calorimetry. Therefore, we conclude that tracking detectors at a Higgs factory could still be capable of providing excellent sensitivity to new light scalars beyond the Standard Model coupling to the $Z$. 

\section{Discussion}
\label{sec:discussion}

A high-precision determination of the inclusive $Zh$ cross section with the recoil method is among the most important physics targets for future $e^+e^-$ Higgs factories. It is of academic and possibly practical interest to know what minimum detector elements are required to perform this measurement. In this work we have shown that a measurement of $\sigma_{Zh}$ in the $Z\rightarrow\mu^+\mu^-$ channel using only tracking information can achieve nearly the same precision as the conventional measurement utilizing full calorimetry and tracking. At 250 GeV with 200 fb$^{-1}$ (achievable within a year or two of running), the track-based analysis can reach $\delta\sigma_{Zh}/\sigma_{Zh}\approx 6\%$, compared with $\delta\sigma_{Zh}/\sigma_{Zh}\approx 5\%$ in the full detector analysis. The primary advantage of calorimetry in this analysis is the ability to identify photons. Recovering photons radiated from final state muons increases the signal efficiency by about 10\%, while rejecting events with hard ISR photons results in a background reduction of $\approx 40\%$. A tracker-only detector might also be supplemented by a simple photon veto conversion layer to achieve the same goal, but in any case the improvement in the expected sensitivity is small in this analysis.

The $Z\rightarrow e^+e^-$ channel is more challenging due to an increase in backgrounds, larger bremsstrahlung effects, and the difficulty of distinguishing electrons from charged hadrons without calorimetry. We have not performed a complete study of this channel, but we have shown that $e^\pm$ and $\pi^\pm$ can be efficiently discriminated in a tracker by isolation requirements.

Finally, in the $\mu^+\mu^-+X$ channel one can also perform a bump-hunt in $m_{\rm \tiny{reco}}$, providing a decay-independent search for new light scalars that couple to the $Z$, for example through mixing with the Higgs. The tracker-only search offers comparable sensitivity for $m_\phi\gtrsim m_W$ and sensitivity within a factor of a few of the full detector analysis at lower masses. The latter is again controlled by background involving ISR photons and the track-only search could be improved by the photon veto conversion layer.

Our results suggest several avenues for future work. It would be of interest to perform a full study of the $e^+e^-$ channel, as well as exclusive channels. $Zh\rightarrow bbjj$ is an interesting case where there is enough kinematic information (together with the $Z$ and $h$ masses) to reconstruct the event, even if neutral hadronic energy goes unmeasured. It would also be worthwhile to consider modifications of other detector subsystems and their impact on the science reach at Higgs factories. We suspect that further study along these lines may reveal new practically and economically advantageous approaches to studying the Higgs boson at next-generation lepton colliders.

\section*{Acknowledgements}

We would like to thank John Paul Chou, Nathaniel Craig, Michael Dine, Yuri Gershtein, Joe Incandela, and Sunil Somalwar
for useful discussions and encouragement, and the Kavli Institute for Theoretical Physics 
where this work was begun.  
The work of PD and JK was supported by NSF grant PHY-1719642. The work of ST was supported 
by the US Department of Energy under grant DE-SC0010008, 
and in part by the National Science Foundation under Grant NSF PHY-1748958.  JK gratefully acknowledges the hospitality of the Aspen Center for Physics, supported by National Science Foundation grant PHY-1607611, where a portion of this work was completed.

\bibliography{Precision_Higgs_tracking}{}
\bibliographystyle{utphys}

\end{document}